\documentclass[12pt,a4paper]{article}
\usepackage[cp850]{inputenc}
\usepackage{amssymb}
\usepackage{latexsym}
\usepackage[dvips]{epsfig}
\usepackage{pstricks,pst-node}
\renewcommand{\baselinestretch}{1.5}
\begin{document}
\begin{titlepage} 
\renewcommand{\baselinestretch}{1}
\small\normalsize
\begin{flushright}
hep-th/0110021\\
MZ-TH/01-26     \\
\end{flushright}

\vspace{1cm}

\begin{center}   

{\LARGE \sc Is Quantum Einstein Gravity\\[5mm]   
Nonperturbatively Renormalizable?}

\vspace{2cm}
{\large O. Lauscher and M. Reuter}\\

\vspace{1cm}
\noindent
{\it Institute of Physics, University of Mainz\\
Staudingerweg 7, D-55099 Mainz, Germany}\\

\end{center}   
 
\vspace*{1.5cm}
\begin{abstract}
We find considerable evidence supporting the conjecture that four-dimensional 
Quantum Einstein Gravity is ``asymptotically safe'' in Weinberg's sense. This
would mean that the theory is likely to be nonperturbatively renormalizable
and thus could be considered a fundamental (rather than merely effective)
theory which is mathematically consistent and predictive down to arbitrarily 
small length scales. For a truncated version of the exact flow equation of the
effective average action we establish the existence of a non-Gaussian 
renormalization group fixed point which is suitable for the construction of a 
nonperturbative infinite cutoff-limit. The truncation ansatz includes the 
Einstein-Hilbert action and a higher derivative term.
\end{abstract}
\end{titlepage}
%
Classical General Relativity, based upon the Einstein-Hilbert action
\begin{eqnarray}
\label{l1}
S_{\rm EH}=\frac{1}{16\pi G}\int d^4x\,\sqrt{-g}\,\left\{-R+2\Lambda
\right\}\;,
\end{eqnarray}
is known to be a phenomenologically very successful theory at length scales
ranging from terrestrial scales to solar system and cosmological
scales. However, it is also known that quantized General Relativity is
perturbatively nonrenormalizable. This has led to the widespread believe that
a straightforward quantization of the metric degrees of freedom cannot lead to
a mathematically consistent and predictive {\it fundamental} theory valid down
to arbitrarily small spacetime distances. Einstein gravity is rather
considered merely an {\it effective} theory whose range of applicability is
limited to a phenomenological description of gravitational effects at
distances much larger than the Planck length.

In particle physics one usually considers a theory fundamental if it is
perturbatively renormalizable. The virtue of such models is that one can
``hide'' their infinities in only finitely many basic parameters (masses,
gauge couplings, etc.) which are intrinsically undetermined within the theory
and whose value must be taken from the experiment. All higher
couplings are then well-defined computable functions of those few parameters.
In nonrenormalizable effective theories, on the other hand, the divergence
structure is such that increasing orders of the loop expansion require an
increasing number of new counter terms and, as a consequence, of undetermined
free parameters. Typically, at high energies, all these unknown parameters
enter on an equal footing so that the theory looses all its predictive power.

However, there are examples of field theories which ``exist'' as fundamental
theories despite their perturbative nonrenormalizability \cite{wilson,parisi}.
These models are ``nonperturbatively renormalizable'' along the lines of
Wilson's modern formulation of renormalization theory \cite{wilson}. They are
constructed by performing the limit of infinite ultraviolet cutoff
(``continuum limit'') at a non-Gaussian renormalization group fixed point
${\rm g}_{*i}$ in the space $\{{\rm g}_i\}$ of all (dimensionless, essential) 
couplings ${\rm g}_i$ which parametrize a general action functional. This has 
to be contrasted with the standard perturbative renormalization which, at least
implicitly, is based upon the Gaussian fixed point at which all couplings
vanish, ${\rm g}_{*i}=0$ \cite{polch,bag}.

In his ``asymptotic safety'' scenario Weinberg \cite{wein,wein2} has put
forward the idea that perhaps a quantum field theory of gravity can be
constructed nonperturbatively by invoking a non-Gaussian ultraviolet (UV)
fixed point $({\rm g}_{*i}\neq 0)$. The resulting theory would be 
``asymptotically
safe'' in the sense that at high energies unphysical singularities are likely
to be absent. To sketch the basic idea let us define the UV critical surface
${\cal S}_{\rm UV}$ to consist of all renormalization group (RG) trajectories
hitting the non-Gaussian fixed point in the infinite cutoff limit. Its
dimensionality ${\rm dim}\left({\cal S}_{\rm UV}\right)\equiv \Delta_{\rm UV}$
is given by the number of attractive (for {\it in}creasing cutoff $k$) 
directions in the space of couplings. Writing the RG equations as
\begin{eqnarray}
\label{l2}
k\,\partial_k {\rm g}_i=\mbox{\boldmath $\beta$}_i({\rm g}_1,{\rm g}_2,\cdots)
\;,
\end{eqnarray}
the linearized flow near the fixed point is governed by the Jacobi matrix
${\bf B}=(B_{ij})$, $B_{ij}\equiv\partial_j
\mbox{\boldmath$\beta$}_i({\rm g}_*)$:
\begin{eqnarray}
\label{p2.1}
k\,\partial_k\,{\rm g}_i(k)=\sum\limits_j B_{ij}\,\left({\rm g}_j(k)
-{\rm g}_{*j}\right)\;.
\end{eqnarray}
The general solution to this equation reads
\begin{eqnarray}
\label{p2.8}
{\rm g}_i(k)={\rm g}_{*i}+\sum\limits_I C_I\,V^I_i\,
\left(\frac{k_0}{k}\right)^{\theta_I}
\end{eqnarray}
where the $V^I$'s are the right-eigenvectors of ${\bf B}$ with eigenvalues 
$-\theta_I$, i.e. $\sum_{j} B_{ij}\,V^I_j=-\theta_I\,V^I_i$. Since ${\bf B}$ is
not symmetric in general the $\theta_I$'s are not guaranteed to be real. We
assume that the eigenvectors form a complete system though. Furthermore, $k_0$ 
is a fixed reference scale, and the $C_I$'s are constants of integration. If
${\rm g}_i(k)$ is to approach ${\rm g}_{*i}$ in the infinite cutoff limit
$k\rightarrow\infty$ we must set $C_I=0$ for all $I$ with 
${\rm Re}\,\theta_I<0$. Hence the dimensionality $\Delta_{\rm UV}$ equals the 
number of ${\bf B}$-eigenvalues with a negative real part, i.e. the number of
$\theta_I$'s with a positive real part.

When we send the cutoff to infinity we must pick one of the trajectories on 
${\cal S}_{\rm UV}$ in order to specify a continuum limit. This means that
we have to fix $\Delta_{\rm UV}$ free parameters, which are not predicted
by the theory and, in principle, should be taken from experiment. Stated
differently, when we {\it lower} the cutoff, only $\Delta_{\rm UV}$ parameters
in the initial action are ``relevant'' and must be fine-tuned in order to place
the system on ${\cal S}_{\rm UV}$; the remaining ``irrelevant'' parameters are
all attracted towards ${\cal S}_{\rm UV}$ automatically. Therefore the theory
has the more predictive power the smaller is the dimensionality of
${\cal S}_{\rm UV}$, i.e. the fewer UV attractive eigendirections the
non-Gaussian fixed point has. If $\Delta_{\rm UV}<\infty$, the quantum field 
theory thus constructed is comparable to and as predictive as a perturbatively
renormalizable model with $\Delta_{\rm UV}$ ``renormalizable couplings'', i.e.
couplings relevant at the Gaussian fixed point.

It is plausible that ${\cal S}_{\rm UV}$ is indeed finite dimensional. If the
dimensionless ${\rm g}_i$'s arise as ${\rm g}_i(k)=k^{-d_i}{\rm G}_i(k)$ by 
rescaling (with the cutoff $k$) the original couplings ${\rm G}_i$ with mass 
dimensions $d_i$, then $\mbox{\boldmath $\beta$}_i=-d_i{\rm g}_i+\cdots$ and 
$B_{ij}=-d_i\delta_{ij}+\cdots$ where the dots stand for the loop 
contributions. Ignoring them, $\theta_i=d_i+\cdots$, and $\Delta_{\rm UV}$
equals the number of positive $d_i$'s. Since adding derivatives or powers of
fields to a monomial in the action always lowers $d_i$, there can be at most
a finite number of positive $d_i$'s and, therefore, of negative eigenvalues
of ${\bf B}$. Thus, barring the possibility that the loop corrections change
the signs of infinitely many elements in ${\bf B}$, the dimensionality of
${\cal S}_{\rm UV}$ is finite \cite{wein}.

In the present paper we investigate the fixed point structure of Euclidean
Einstein gravity in the framework of the effective average action 
\cite{avact,avact2} which was developed for quantum gravity in \cite{Reu96}. 
Here the term ``Einstein gravity'' stands for the class of theories which are
described by action functionals depending on the metric as the dynamical
variable and which respect general coordinate invariance. 

The effective average action $\Gamma_k$ is a coarse grained free energy
functional that describes the behavior of the theory at the mass scale $k$ 
\cite{avact}. It contains the quantum effects of all fluctuations of the 
dynamical variable with momenta larger than $k$, but not of those 
with momenta smaller than $k$. As $k$ is decreased, an increasing number 
of degrees of freedom is integrated out. This successive averaging of the
fluctuation variable is achieved by a $k$-dependent infrared (IR) cutoff term 
$\Delta_kS$ which is added to the classical action in the standard Euclidean 
functional integral. This term gives a momentum dependent mass square ${\cal
R}_k(p^2)$ to the field modes with momentum $p$ which vanishes if $p^2\gg k^2$.
When regarded as a function of $k$, $\Gamma_k$ runs along a RG trajectory in
the space of all action functionals that interpolates between the classical
action $S=\Gamma_{k\rightarrow\infty}$ and the conventional effective action
$\Gamma=\Gamma_{k=0}$. The change of $\Gamma_k$ induced by an infinitesimal
change of $k$ is described by a functional differential equation, the exact
RG equation. In the simplest case it assumes the form
\begin{eqnarray}
\label{newRG}
k\,\partial_k\Gamma_k=\frac{1}{2}\;{\rm Tr}\left[\left(\Gamma_k^{(2)}
+{\cal R}_k\right)^{-1}\,k\,\partial_k{\cal R}_k\right]\;.
\end{eqnarray}
For its derivation \cite{avact,Reu96,tif} and applications 
\cite{avact2,RW97,sven,DP97,odintsov} we refer to the literature. 

In general it is impossible to find an exact solution to eq.(\ref{newRG})
since it describes trajectories in an infinite dimensional space of action
functionals. Hence we are forced to rely upon approximations. A powerful
nonperturbative approximation scheme is the truncation of the ``theory space''
where the RG flow is projected onto a finite-dimensional subspace of the space
of all action functionals. In practice one makes an ansatz for $\Gamma_k$ that
comprises only a few couplings and inserts it into the RG equation. This
leads to a finite set of coupled differential equations of the form
(\ref{l2}). In the following we will use this approach in the context of
quantum gravity.

As a first step we study the fixed point properties of the 
``Einstein-Hilbert truncation'' \cite{Reu96,LR1} defined by the ansatz
\begin{eqnarray}
\label{in2}
\Gamma_k[g,\bar{g}]=\left(16\pi G_k\right)^{-1}\int d^dx\,\sqrt{g}\left\{
-R(g)+2\bar{\lambda}_k\right\}+\mbox{classical gauge fixing}
\end{eqnarray}
which involves only the cosmological constant $\bar{\lambda}_k$ and the Newton
constant $G_k$ as running parameters. All remaining coupling constants are set
to zero in (\ref{in2}). The gauge fixing term in (\ref{in2}) contains the 
parameter $\alpha$ whose evolution is neglected. The value $\alpha=0$ is of
special importance because it can be argued to be a RG fixed point 
\cite{alpha}. As a consequence, setting $\alpha=0$ by hand mimics the
dynamical treatment of the gauge fixing parameter. Inserting (\ref{in2}) into
the gravitational analog of the RG equation (\ref{newRG}) one obtains a set of 
$\mbox{\boldmath$\beta$}$-functions $(\mbox{\boldmath$\beta$}_\lambda,
\mbox{\boldmath$\beta$}_g)$ for the dimensionless cosmological constant 
$\lambda_k\equiv k^{-2}\bar{\lambda}_k$ and the dimensionless Newton constant
$g_k\equiv k^{d-2}G_k$, respectively. They describe a two-dimensional RG flow 
on the plane with coordinates ${\rm g}_1\equiv\lambda$ and ${\rm g}_2\equiv g$.

The fixed points of the RG flow are precisely those points in the space of
dimensionless, essential couplings where all 
$\mbox{\boldmath$\beta$}$-functions
vanish simultaneously. In the context of the Einstein-Hilbert truncation there
exist both a trivial Gaussian fixed point at $\lambda_*=g_*=0$ and a 
UV attractive non-Gaussian fixed point at $(\lambda_*,g_*)\neq(0,0)$.

An immediate consequence of the non-Gaussian fixed point is that, for
$k\rightarrow\infty$,
\begin{eqnarray}
\label{p2.9}
G_k\approx \frac{g_*}{k^{d-2}}\;\;,\;\;\;\;\;\bar{\lambda}_k\approx\lambda_*\,
k^2\;.
\end{eqnarray}
The vanishing of Newton's constant at high momenta means that gravity is
{\it asymptotically free} (for $d>2$).

The case of $d=2+\varepsilon$ dimensions $(0<|\varepsilon|\ll 1)$ has been 
widely discussed in the literature \cite{wein,Reu96,gastmans}. Within the 
average action approach, one finds that the non-Gaussian fixed
point is located at
\begin{eqnarray}
\label{p2.2}
\lambda_*(\varepsilon)=-\frac{3}{38}\Phi^1_1(0)\,\varepsilon+{\cal O}\left(
\varepsilon^2\right)\;,\;\;\;
g_*(\varepsilon)=\frac{3}{38}\,\varepsilon+{\cal O}\left(
\varepsilon^2\right)
\end{eqnarray}
where $\Phi^1_1(0)$ is a cutoff scheme dependent parameter of order unity 
\cite{gastmans}.
Linearizing the flow equation in the vicinity of the fixed point \cite{LR1} we
obtain the critical exponents $\theta_1=2-\frac{12\alpha-13}{19}\,
\varepsilon+{\cal O}(\varepsilon^2)$ and $\theta_2=\varepsilon
+{\cal O}(\varepsilon^2)$ as the eigenvalues of $-{\bf B}$. Both critical 
exponents are positive for $\varepsilon>0$ and cutoff scheme independent up to
terms of ${\cal O}(\varepsilon^2)$. Hence the non-Gaussian fixed point is UV 
attractive in both directions of the $\lambda$-$g-$plane for $\varepsilon>0$.
All trajectories in its basin of attraction run into the fixed point for 
$k\rightarrow\infty$.

In the most interesting case of $d=4$ dimensions the non-Gaussian fixed point 
was first analyzed, within this framework, in \cite{souma1,bh,souma2}, and its 
possible 
implications for black hole physics \cite{bh} and cosmology \cite{cosmo1} were
studied. The average action approach used in these investigations is 
nonperturbative in nature and does not rely upon the $\varepsilon$-expansion.

The question of crucial importance is whether the fixed point predicted by
the Einstein-Hilbert truncation actually approximates a fixed point in the
exact theory, or whether it is an artifact of the truncation. In this letter
we summarize the evidence we found recently \cite{LR1,LR3} which in our opinion
strongly supports the hypothesis that there indeed exists a non-Gaussian fixed
point in the exact 4-dimensional theory, with exactly the properties required
by the asymptotic safety scenario.

We have tried to assess the reliability of the Einstein-Hilbert truncation
both by analyzing the cutoff scheme dependence within this truncation 
\cite{LR1} and by generalizing the truncation ansatz \cite{LR3}.

The cutoff operator ${\cal R}_k(p^2)$ is specified by a matrix in field space
and a ``shape function'' $R^{(0)}(p^2/k^2)$ which describes the details of how
the modes get suppressed in the IR when $p^2$ drops below $k^2$. As for the
matrix in field space, a cutoff of ``type A'' was used in the original paper
\cite{Reu96}, while a new cutoff of ``type B'' was constructed in \cite{LR1}.
The latter is natural and convenient from a technical point of view when one
uses the transverse-traceless decomposition of the metric. As for the shape
function, we employed the one-parameter family of exponential cutoffs 
\cite{souma2}
\begin{eqnarray}
\label{p2.7}
R^{(0)}(y)=sy\left[\exp(sy)-1\right]^{-1}
\end{eqnarray}
as well as a similar family of functions with compact support. In \cite{frank}
also a sharp cutoff was used. In (\ref{p2.7}) the ``shape parameter'' $s$ 
allows us to change the profile of $R^{(0)}$. We checked the cutoff scheme
dependence of the various quantities of interest both by looking at their
dependence on $s$ and similar shape parameters, and by comparing the ``type A''
and ``type B'' results.

Universal quantities are particularly important because, by definition, they
are strictly cutoff scheme independent in the exact theory. Any truncation
leads to a scheme dependence of these quantities, however, and its magnitude
is a natural indicator for the quality of the truncation \cite{kana}. Typical
examples of universal quantities are the critical exponents $\theta_I$. The
existence or nonexistence of a fixed point is also a universal, scheme 
independent feature, but its precise location in parameter space is scheme
dependent. Nevertheless it can be argued that, in $d=4$, the product $g_*
\lambda_*$ is universal \cite{LR1} while $g_*$ and $\lambda_*$ separately are
not.

The ultimate justification of a certain truncation is that when one adds 
further terms to it its physical predictions do not change significantly any 
more. As a first step towards testing the stability of the Einstein-Hilbert
truncation against the inclusion of other invariants we took a $({\rm 
curvature})^2$-term into account:
\begin{eqnarray}
\label{p2.4}
\Gamma_k[g,\bar{g}]=\int d^dx\,\sqrt{g}\left\{\left(16\pi G_k\right)^{-1}
\left[-R(g)+2\bar{\lambda}_k\right]+\bar{\beta}_k\,R^2(g)\right\}
+\mbox{classical gauge fixing}.
\end{eqnarray}
Inserting (\ref{p2.4}) into the functional RG equation yields a set of 
$\mbox{\boldmath$\beta$}$-functions 
$(\mbox{\boldmath$\beta$}_\lambda,\mbox{\boldmath$\beta$}_g,
\mbox{\boldmath$\beta$}_\beta)$ for the dimensionless couplings $\lambda_k$,
$g_k$ and $\beta_k\equiv k^{4-d}\bar{\beta}_k$. They describe the RG flow on
the three-dimensional $\lambda$-$g$-$\beta-$space. In order to make the
technically rather involved calculation of the
$\mbox{\boldmath$\beta$}$-functions feasible we had to restrict ourselves to
the gauge $\alpha=1$.

Beyond $R^2$ there exist two more $({\rm curvature})^2$-terms: $R_{\mu\nu}
R^{\mu\nu}$ and $R_{\mu\nu\rho\sigma}R^{\mu\nu\rho\sigma}$. We omitted the
latter two terms from the generalized truncation for both a technical and a
conceptual reason: (i) In order to project the RG flow onto the truncation
subspace we inserted the metrics for a family of spheres $S^d$, 
parametrized by their radius $r$, into the flow equation. While this
projection technique is capable of distinguishing $\int d^dx\,\sqrt{g}\propto 
r^{d}$ from $\int d^dx\,\sqrt{g}R\propto r^{d-2}$, it cannot disentangle
$\int d^dx\,\sqrt{g}R^2$, $\int d^dx\,\sqrt{g}R_{\mu\nu}^2$, and 
$\int d^dx\,\sqrt{g}R_{\mu\nu\rho\sigma}^2$ which are all proportional to
$r^{d-4}$. If one wants to project out the three
$({\rm curvature})^2$-invariants individually one must insert spaces which are
not maximally symmetric, but then the evaluation of the pertinent functional
traces is a rather formidable problem with the present technology. (ii) Doing
perturbation theory about the Gaussian fixed point, consistency requires to
include either no $({\rm curvature})^2$-term at all, or all three of them, the
reason being that they have the same canonical dimension and are equally 
``relevant'' therefore. However, the non-Gaussian fixed point has its own
scaling fields and dimensions according to which the different
$({\rm curvature})^2$-terms are not on an equal footing anyhow.

In the following we summarize the pieces of evidence pointing in the direction
that the exact Quantum Einstein Gravity might indeed be ``asymptotically safe''
in 4 dimensions. For details we refer to \cite{LR1} and \cite{LR3}. We begin
by listing the results obtained with the pure Einstein-Hilbert truncation
without the $R^2$-term:

{\bf (1)} \underline{Universal Existence}: 
Both for type A and type B cutoffs the non-Gaussian fixed point exists for all
shape functions $R^{(0)}$ we considered. (This generalizes earlier results in
\cite{souma1,souma2}.) It seems impossible to find an admissible cutoff which
destroys the fixed point in $d=4$. This result is highly nontrivial since in
higher dimensions $(d\gtrsim 5)$ the fixed point exists for some but does not
exist for other cutoffs \cite{frank}.

{\bf (2)} \underline{Positive Newton Constant}:
While the position of the fixed point is scheme dependent, all cutoffs yield
{\it positive} values of $g_*$ and $\lambda_*$. A negative $g_*$ might be
problematic for stability reasons, but there is no mechanism in the flow
equation which would exclude it on general grounds. In fact, eq.(\ref{p2.2})
shows that $g_*<0$ for $d<2$.

{\bf (3)} \underline{Stability}:
For any cutoff employed the non-Gaussian fixed point is found to be UV
attractive in both directions of the $\lambda$-$g-$plane. Linearizing the
flow equation according to eq. (\ref{p2.1}) we obtain a pair of complex
conjugate critical exponents $\theta_1=\theta_2^*$ with positive real part 
$\theta'$ and imaginary parts $\pm\theta''$. Introducing $t\equiv\ln(k/k_0)$
the general solution to the linearized flow equations reads
\begin{eqnarray}
\label{p2.5}
\left(\lambda_k,g_k\right)^{\bf T}
&=&\left(\lambda_*,g_*\right)^{\bf T}
+2\Bigg\{\left[{\rm Re}\,C\,\cos\left(\theta''\,t\right)
+{\rm Im}\,C\,\sin\left(\theta''\,t\right)\right]
{\rm Re}\,V\nonumber\\
& &+\left[{\rm Re}\,C\,\sin\left(\theta''\,t\right)-{\rm Im}\,C
\,\cos\left(\theta''\,t\right)\right]{\rm Im}\,V\Bigg\}e^{-\theta' t}\;.
\end{eqnarray}
with $C\equiv C_1=(C_2)^*$ an arbitrary complex number and $V\equiv 
V^1=(V^{2})^*$ the right-eigenvector of ${\bf B}$ with eigenvalue $-\theta_1
=-\theta_2^*$. Eq.(\ref{p2.5}) implies that, due to the positivity of 
$\theta'$, all trajectories hit the fixed point as $t$ is sent to infinity. 
The nonvanishing imaginary part $\theta''$ has no impact on the stability. 
However, it influences the shape of the trajectories which spiral into the 
fixed point for $k\rightarrow\infty$. Thus, also in 4 dimensions the fixed 
point has the stability properties expected in the asymptotic safety scenario.

Solving the full, nonlinear flow equations \cite{frank} shows that the 
asymptotic scaling region where the linearization (\ref{p2.5}) is valid 
extends from $k``=\mbox{''}\infty$ down to about $k\approx m_{\rm Pl}$ with the
Planck mass defined as $m_{\rm Pl}\equiv G_0^{-1/2}$. (It plays a role similar
to $\Lambda_{\rm QCD}$ in QCD.) It is the regime above the Planck scale where
gravity becomes weakly coupled and asymptotically free. We set $k_0\equiv
m_{\rm Pl}$ so that the asymptotic scaling regime extends from about $t=0$ to
$t``=\mbox{''}\infty$.

{\bf (4)} \underline{Scheme- and Gauge Dependence}:
We analyzed the cutoff scheme dependence of $\theta'$, $\theta''$, and 
$g_*\lambda_*$ as a measure for the reliability of the truncation.
The critical exponents were found to be reasonably constant within about a 
factor of 2. For $\alpha=1$ and
$\alpha=0$, for instance, they assume values in the ranges $1.4\lesssim
\theta'\lesssim 1.8$, $2.3\lesssim\theta''\lesssim 4$ and $1.7\lesssim
\theta'\lesssim 2.1$, $2.5\lesssim\theta''\lesssim 5$, respectively. The
universality properties of the product $g_*\lambda_*$ are much more
impressive though. Despite the rather strong scheme dependence of $g_*$ and 
$\lambda_*$ separately, their product has almost no visible $s$-dependence for
not too small values of $s$. Its value is
\begin{eqnarray}
\label{product}
g_*\lambda_*\approx\left\{\begin{array}{l}\mbox{$0.12$ for $\alpha=1$}\\
\mbox{$0.14$ for $\alpha=0$}\end{array}\right.\;.
\end{eqnarray}
The differences between the ``physical'' (fixed point) value of the gauge
parameter, $\alpha=0$, and the technically more convenient $\alpha=1$ are at 
the level of about 10 to 20 per-cent. 

Up to this point all results referred to the pure Einstein-Hilbert truncation
(\ref{in2}). Next we list the main results obtained with the generalized
truncation (\ref{p2.4}) which includes the $R^2$-term.

\begin{figure}[ht]
\begin{minipage}{7.9cm}
        \epsfxsize=7.9cm
        \epsfysize=5.2cm
        \centerline{\epsffile{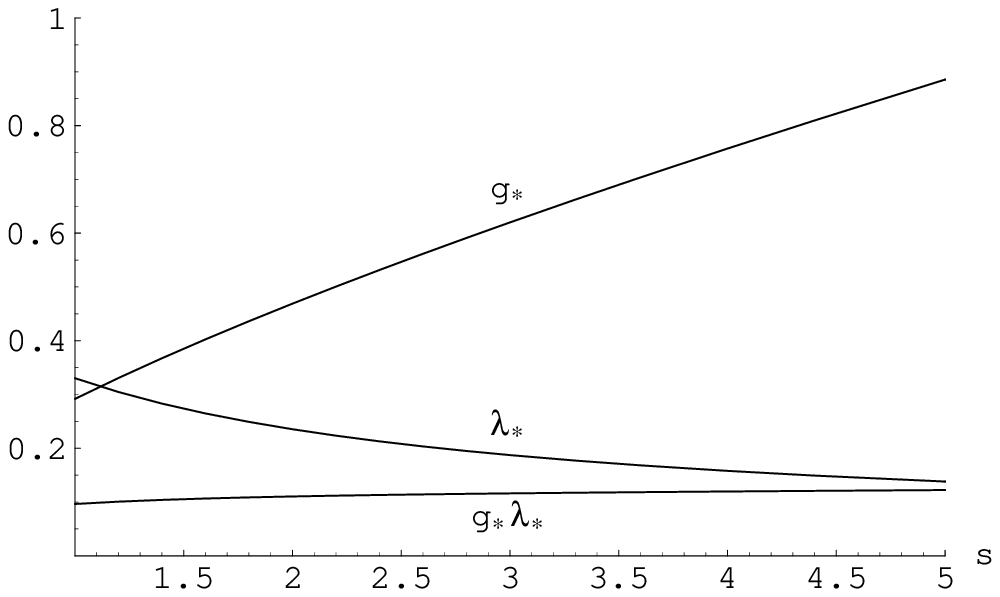}}
\centerline{(a)}
\end{minipage}
\hfill
\begin{minipage}{7.5cm}
        \epsfxsize=7.5cm
        \epsfysize=5cm
        \centerline{\epsffile{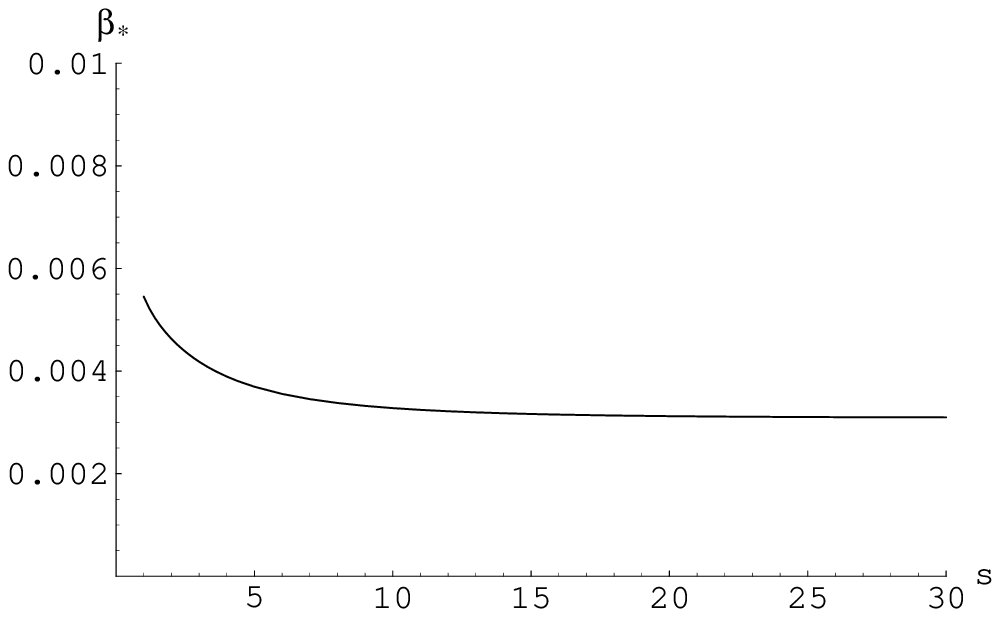}}
\centerline{(b)}
\end{minipage}
\vspace{0.3cm}
\caption{(a) $g_*$, $\lambda_*$, and $g_*\lambda_*$ as functions of $s$ 
for $1\le s\le 5$, and (b) $\beta_*$ as a function of $s$ for $1\le s\le 30$,
using the family of exponential shape functions (\ref{p2.7}).}  
\label{plot1}
\end{figure}

{\bf (5)} \underline{Position of the Fixed Point $(R^2)$}:
Also with the generalized truncation the fixed point is found to exist for all
admissible cutoffs. In FIG. \ref{plot1} we show its coordinates 
$(\lambda_*,g_*,\beta_*)$ for the shape functions (\ref{p2.7}) and
the type B cutoff. For every shape parameter $s$, the values of $\lambda_*$ 
and $g_*$ are almost the same as those obtained with the Einstein-Hilbert 
truncation. In particular, the product $g_*\lambda_*$ is constant with a very
high accuracy. For $s=1$, for instance, we obtain 
$(\lambda_*,g_*)=(0.348,0.272)$ from the Einstein-Hilbert truncation and
$(\lambda_*,g_*,\beta_*)=(0.330,0.292,0.005)$ from the generalized truncation.
It is quite remarkable that $\beta_*$ is always significantly
smaller than $\lambda_*$ and $g_*$. Within the limited precision of our
calculation this means that in the three-dimensional parameter space the fixed
point practically lies on the $\lambda$-$g-$plane with $\beta=0$, i.e. on the
parameter space of the pure Einstein-Hilbert truncation.

\renewcommand{\baselinestretch}{1}
\small\normalsize
\renewcommand{\baselinestretch}{1.5}
\small\normalsize

{\bf (6)} \underline{Eigenvalues and -vectors $(R^2)$}:
The non-Gaussian fixed point of the
$R^2$-truncation proves to be UV attractive in any of the three directions of
the $\lambda$-$g$-$\beta-$space for all cutoffs used. The linearized flow in
its vicinity is always governed by a pair of complex conjugate critical
exponents $\theta_1=\theta'+{\rm i}\theta''=\theta_2^*$ with $\theta'>0$ and
a single real, positive critical exponent $\theta_3>0$. It may be expressed as
\begin{eqnarray}
\label{p2.6}
\left(\lambda_k,g_k,\beta_k\right)^{\bf T}
&=&\left(\lambda_*,g_*,\beta_*\right)^{\bf T}
+2\Bigg\{\left[{\rm Re}\,C\,\cos\left(\theta''\,t\right)
+{\rm Im}\,C\,\sin\left(\theta''\,t\right)\right]
{\rm Re}\,V\nonumber\\
& &+\left[{\rm Re}\,C\,\sin\left(\theta''\,t\right)-{\rm Im}\,C
\,\cos\left(\theta''\,t\right)\right]{\rm Im}\,V\Bigg\}\,e^{-\theta' t}
+C_3 V^3\,e^{-\theta_3 t}
\end{eqnarray}
with arbitrary complex $C\equiv C_1=(C_2)^*$ and arbitrary real $C_3$, and
with $V\equiv V^1=(V^2)^*$ and $V^3$ the right-eigenvectors of the stability
matrix $(B_{ij})_{i,j\in\{\lambda,g,\beta\}}$ with eigenvalues $-\theta_1=
-\theta_2^*$ and $-\theta_3$, respectively. Clearly the conditions for UV 
stability are $\theta'>0$ and $\theta_3>0$. They are indeed satisfied for all 
cutoffs. For the exponential shape function with $s=1$, for instance, we find
$\theta'=2.15$, $\theta''=3.79$, $\theta_3=28.8$, and 
${\rm Re}\,V=(-0.164,0.753,-0.008)^{\bf T}$,
${\rm Im}\,V=(0.64,0,-0.01)^{\bf T}$, $V^3=-(0.92,0.39,0.04)^{\bf T}$. (The
vectors are normalized such that $\left\|V\right\|=\left\|V^3\right\|=1$.)
The trajectories (\ref{p2.6}) comprise three independent normal modes with
amplitudes proportional to ${\rm Re}\,C$, ${\rm Im}C$ and $C_3$, respectively.
The first two are of the spiral type again, the third one is a straight line.

For any cutoff, the numerical results have several quite remarkable
properties. They all indicate that, close to the non-Gaussian fixed point, the
RG flow is rather well approximated by the pure Einstein-Hilbert truncation.

{\bf (a)}
The $\beta$-components of ${\rm Re}\,V$ and ${\rm Im}\,V$ are very
tiny. Hence these two vectors span a plane which virtually coincides with the
$\lambda$-$g-$subspace at $\beta=0$, i.e. with the parameter space of the
Einstein-Hilbert truncation. As a consequence, the  ${\rm Re}\,C$- and
${\rm Im}C$- normal modes are essentially the same trajectories as the
``old'' normal modes already found without the $R^2$-term. Also the
corresponding $\theta'$- and $\theta''$-values coincide within the scheme
dependence.

{\bf (b)}
The new eigenvalue $\theta_3$ introduced by the $R^2$-term is
significantly larger than $\theta'$. When a trajectory approaches the fixed
point from below $(t\rightarrow\infty)$, the ``old'' normal modes $\propto
{\rm Re}\,C,{\rm Im}\,C$ are proportional to $\exp(-\theta' t)$, but the new
one is proportional to $\exp(-\theta_3 t)$, so that it decays much more
quickly. For every trajectory running into the fixed point, i.e. for every
set of constants $({\rm Re}\,C,{\rm Im}\,C,C_3)$, we find therefore that once
$t$ is sufficiently large the trajectory lies entirely in the
${\rm Re}\,V$-${\rm Im}\,V-$subspace, i.e. the $\beta=0$-plane practically.

Due to the large value of $\theta_3$, the new scaling field is very
``relevant''. However, when we start at the fixed point $(t``=\mbox{''}\infty)$
and lower $t$ it is only at the low energy (!) scale $k\approx m_{\rm Pl}$
$(t\approx 0)$ that $\exp(-\theta_3 t)$ reaches unity, and only then, i.e.
far away from the fixed point, the new scaling field starts growing
rapidly.

{\bf (c)}
Since the matrix ${\bf B}$ is not symmetric its eigenvectors have
no reason to be orthogonal. In fact, we find that $V^3$ lies almost in the
${\rm Re}\,V$-${\rm Im}\,V-$plane. For the angles between the eigenvectors
given above we obtain $\sphericalangle ({\rm Re}\,V,{\rm Im}\,V)=102.3^\circ$, 
$\sphericalangle({\rm Re}\,V,V^3)=100.7^\circ$, 
$\sphericalangle({\rm Im}\,V,V^3)=156.7^\circ$. Their sum is $359.7^\circ$
which confirms that ${\rm Re}\,V$, ${\rm Im}\,V$ and $V^3$ are almost 
coplanar. This implies that when we lower $t$ and move away from the fixed
point so that the $V^3$- scaling field starts growing, it is again
predominantly the $\int d^dx\,\sqrt{g}$- and $\int d^dx\,\sqrt{g}R$-invariants
which get excited, but not $\int d^dx\,\sqrt{g}R^2$ in the first place.

\begin{figure}[ht]
\begin{minipage}{7.9cm}
        \epsfxsize=7.9cm
        \epsfysize=5.2cm
        \centerline{\epsffile{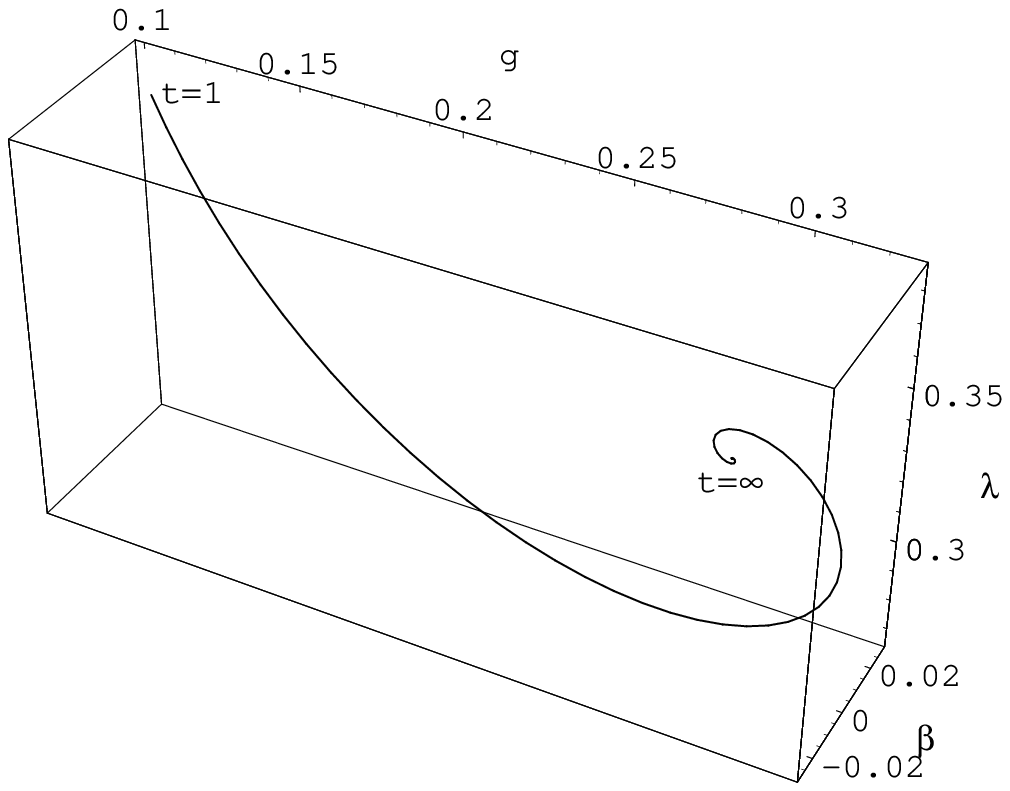}}
\centerline{(a)}
\end{minipage}
\hfill
\begin{minipage}{7.9cm}
        \epsfxsize=7.9cm
        \epsfysize=5.2cm
        \centerline{\epsffile{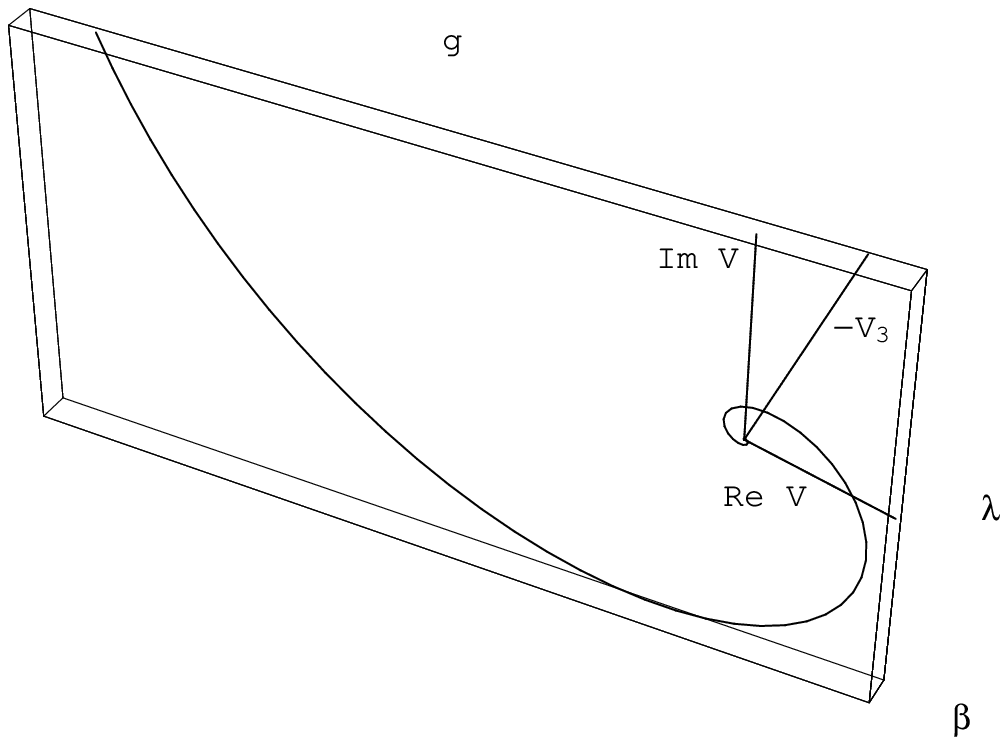}}
\centerline{(b)}
\end{minipage}
\vspace{0.3cm}
\caption{Trajectory of the linearized flow equation obtained from the
$R^2$-truncation for $1\le t=\ln(k/k_0)<\infty$. In (b) we depict
the eigendirections and the ``box'' to which the trajectory is confined.}  
\label{plot2}
\end{figure}

Summarizing the three points above we can say that very close to the fixed
point the RG flow seems to be essentially two-dimensional, and that this
two-dimensional flow is well approximated by the RG equations of the
Einstein-Hilbert truncation. In FIG. \ref{plot2} we show a typical trajectory
which has all three normal modes excited with equal strength
$({\rm Re}\,C={\rm Im}\,C=1/\sqrt{2}$, $C_3=1)$. All its way down from
$k``=\mbox{''}\infty$ to about $k=m_{\rm Pl}$ it is confined to a very thin
box surrounding the $\beta=0$-plane.

\renewcommand{\baselinestretch}{1}
\small\normalsize
\renewcommand{\baselinestretch}{1.5}
\small\normalsize

\begin{figure}[ht]
\begin{minipage}{7.9cm}
        \epsfxsize=7.9cm
        \epsfysize=5.2cm
        \centerline{\epsffile{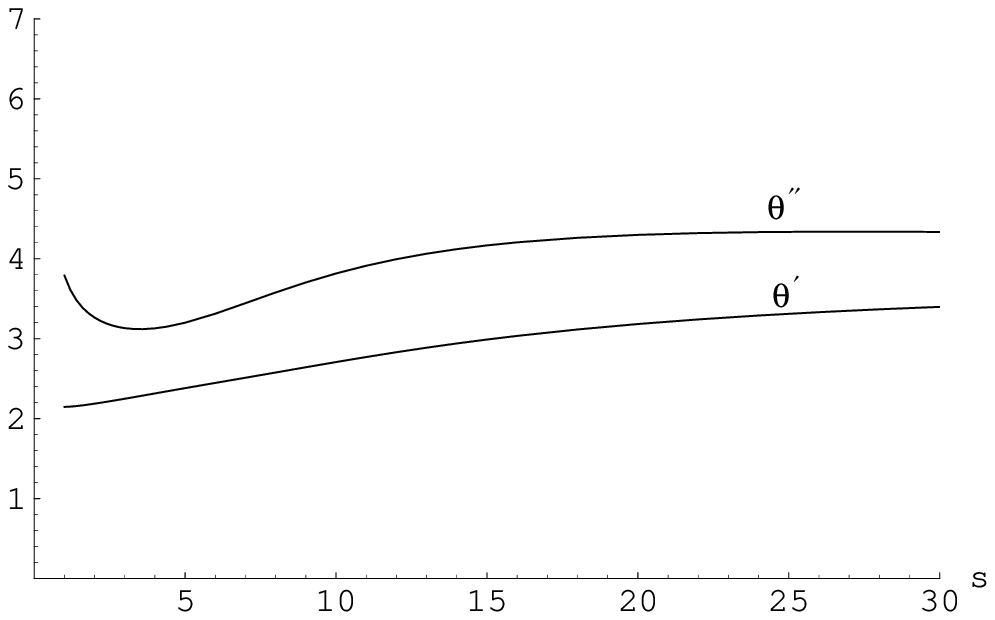}}
\centerline{(a)}
\end{minipage}
\hfill
\begin{minipage}{7.9cm}
        \epsfxsize=7.9cm
        \epsfysize=5.2cm
        \centerline{\epsffile{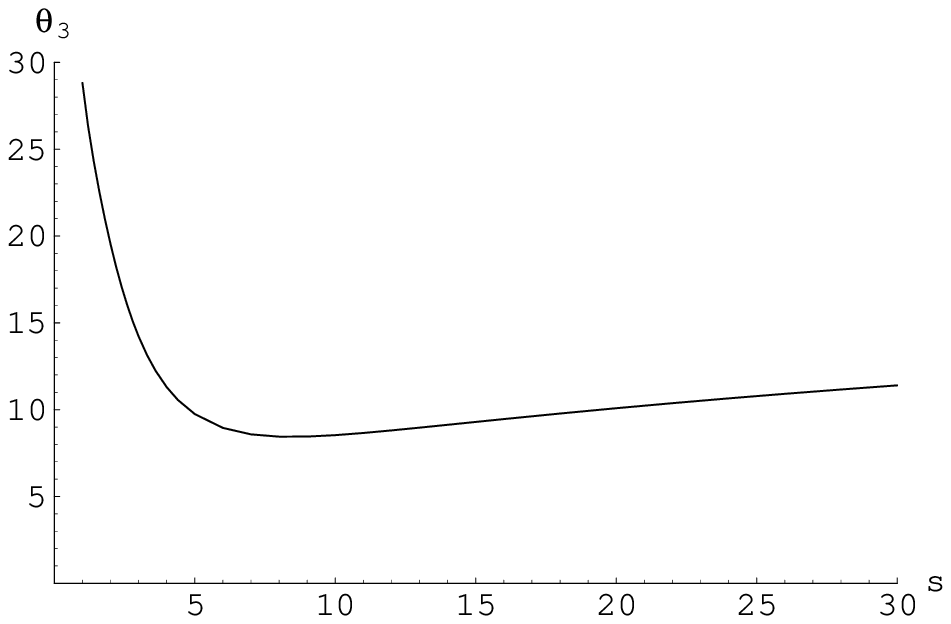}}
\centerline{(b)}
\end{minipage}
\vspace{0.3cm}
\caption{(a) $\theta'={\rm Re}\,\theta_1$ and $\theta''={\rm Im}\,\theta_1$, 
and (b) $\theta_3$ as functions of $s$, using the family of exponential shape 
functions (\ref{p2.7}).}  
\label{plot3}
\end{figure}

{\bf (7)} \underline{Scheme Dependence $(R^2)$}:
The scheme dependence of the critical exponents and of the product
$g_*\lambda_*$ turns out to be of the same order of magnitude as in the case
of the Einstein-Hilbert truncation. FIG. \ref{plot3} shows the cutoff
dependence of the critical exponents, using the family of shape functions
(\ref{p2.7}). For the cutoffs employed $\theta'$ and $\theta''$ assume values
in the ranges $2.1\lesssim\theta'\lesssim 3.4$ and $3.1\lesssim\theta''
\lesssim 4.3$, respectively. While the scheme dependence of $\theta''$ is
weaker than in the case of the Einstein-Hilbert truncation we find that it is
slightly larger for $\theta'$. The exponent $\theta_3$ suffers from relatively
strong variations as the cutoff is changed, $8.4\lesssim\theta_3\lesssim
28.8$, but it is always significantly larger than $\theta'$.
The product $g_*\lambda_*$ again exhibits an extremely weak scheme dependence.
FIG. \ref{plot1}(a) displays $g_*\lambda_*$ as a function of $s$. 
It is impressive to see how the cutoff
dependences of $g_*$ and $\lambda_*$ cancel almost perfectly. FIG.
\ref{plot1}(a) suggests the universal value $g_*\lambda_*\approx 0.14$.
Comparing this value to those obtained from the Einstein-Hilbert truncation,
eq.(\ref{product}), we find that it differs slightly from the one based upon
the same gauge $\alpha=1$. The deviation is of the same size as the difference
between the $\alpha=0$- and the $\alpha=1$-results of the Einstein-Hilbert
truncation. It does not come as a surprise since for technical reasons we did
not use the ``physical'' gauge $\alpha=0$ in the $R^2$-calculation.

\renewcommand{\baselinestretch}{1}
\small\normalsize
\renewcommand{\baselinestretch}{1.5}
\small\normalsize

As for the universality of the critical exponents we emphasize that the
qualitative properties listed above ($\theta',\theta_3>0$, $\theta_3\gg
\theta'$, etc.) obtain universally for all cutoffs. The $\theta$'s have a
much stronger scheme dependence than $g_*\lambda_*$, however. This is most
probably due to neglecting further relevant operators in the truncation so
that the ${\bf B}$-matrix we are diagonalizing is too small still.
Nevertheless it is tempting to compare our figures for the critical exponents
to the numerical simulations of simplicial quantum gravity \cite{amb}. In
\cite{hamber} it was found that $g_k-g_*\propto\exp(-\theta' t)$ with
$\theta'\approx 3$. This value is consistent with our results.

{\bf (8)} \underline{Dimensionality of ${\cal S}_{\rm UV}$}:
According to the canonical dimensional analysis, the
$({\rm cur}$\-va\-${\rm ture})^n$-invariants in 4 dimensions are classically 
marginal for
$n=2$ and irrelevant for $n>2$. The results for $\theta_3$ indicate that there
are large nonclassical contributions so that there might be relevant operators
perhaps even beyond $n=2$. With the present approach it is clearly not
possible to determine their number $\Delta_{\rm UV}$. However, as it is
hardly conceivable that the quantum effects change the signs of arbitrarily
large (negative) classical scaling dimensions, $\Delta_{\rm UV}$ should be
finite. A first confirmation of this picture comes from our $R^2$-calculation
in $d=2+\varepsilon$ where the dimensional count is shifted by two units. In
this case we find indeed that the third scaling field is irrelevant,
$\theta_3<0$. Using the shape function (\ref{p2.7}) with $s=1$, for instance,
we obtain a non-Gaussian fixed point located at $\lambda_*=-0.13\,\varepsilon
+{\cal O}(\varepsilon^2)$, $g_*=0.087\,\varepsilon+{\cal O}(\varepsilon^2)$,
$\beta_*=-0.083+{\cal O}(\varepsilon)$, with critical exponents
$\theta_1=2+{\cal O}(\varepsilon)$,
$\theta_2=0.96\,\varepsilon+{\cal O}(\varepsilon^2)$,
$\theta_3=-1.97+{\cal O}(\varepsilon)$. Therefore the dimensionality of
${\cal S}_{\rm UV}$ could be as small as $\Delta_{\rm UV}=2$, but this is not
a proof, of course. If so, the quantum theory would be characterized by only
two free parameters, the renormalized Newton constant $G_0$ and the
renormalized cosmological constant $\bar{\lambda}_0$.\vspace{24pt}

On the basis of the above results we believe that the non-Gaussian 
fixed
point occuring in the Einstein-Hilbert truncation is very unlikely to be an
artifact of this truncation but rather should be the projection of a fixed
point in the exact theory. We demonstrated explicitly that the fixed point
and all its qualitative properties are stable against variations of the cutoff
and the inclusion of a further invariant in the truncation. It is particularly
remarkable that within the scheme dependence the additional $R^2$-term has
essentially no impact on the fixed point. We interpret the above results and
their mutual consistency as quite nontrivial indications supporting the
conjecture that 4-dimensional Quantum Einstein Gravity indeed possesses a RG
fixed point with precisely the properties needed for its nonperturbative
renormalizability or ``asymptotic safety''. Perhaps quantizing standard
General Relativity does give rise to a microscopic theory of quantum gravity
after all.

\end{document}